# The effect of Auger heating on intraband carrier relaxation in semiconductor quantum rods

MARC ACHERMANN, ANDREW P. BARTKO, JENNIFER A. HOLLINGSWORTH AND VICTOR I. KLIMOV*

Chemistry Division, C-PCS, MS-J567, Los Alamos National Laboratory, Los Alamos, New Mexico 87545, USA

*e-mail: klimov@lanl.gov

The rate at which excited charge carriers relax to their equilibrium state affects many aspects of the performance of nanoscale devices, including switching speed, carrier mobility and luminescent efficiency. Better understanding of the processes that govern carrier relaxation therefore has important technological implications. A significant increase in carrier-carrier interactions caused by strong spatial confinement of electronic excitations in semiconductor nanostructures leads to a considerable enhancement of Auger effects, which can further result in unusual, Auger-process-controlled recombination and energy-relaxation regimes. Here, we report the first experimental observation of efficient Auger heating in CdSe quantum rods at high pump intensities, leading to a strong reduction of carrier cooling rates. In this regime, the carrier temperature is determined by the balance between energy outflow through phonon emission and energy inflow because of Auger heating. This equilibrium results in peculiar carrier cooling dynamics that closely correlate with recombination dynamics, an effect never before seen in bulk or nanoscale semiconductors.

Quantization of electronic and phonon energies and large surface-to-volume ratios significantly modify energy relaxation mechanisms in nanoscale semiconductors compared to bulk materials. In the case of semiconductor nanocrystals[1], strong quantum confinement leads to greatly enhanced carrier-carrier interactions that open new nanocrystal-specific energy relaxation and recombination channels[2-9]. For example, highly efficient electron-hole (e-h) energy transfer (Fig. 1a) can lead to fast, sub-picosecond electron intraband dynamics[3] despite a wide energy separation between quantized states, which can exceed multiple longitudinal optical (LO) phonon energies, $\hbar\omega_0$. Furthermore, enhanced carrier-carrier interactions in nanocrystals result in large rates of nonradiative Auger recombination[6-9], in which the e-h recombination energy is not emitted as a photon but is transferred to an electron or a hole (Fig. 1b). While e-h energy transfer does not change the total energy of the e-h pair, Auger recombination leads to heating of the electronic system (i.e., an increase of the average e-h pair energy) that can, in principle, *slow down* carrier relaxation dynamics.

The role of Auger heating is not significant in bulk semiconductors[10,11] because of the restrictions imposed by energy and momentum conservation. However, Auger processes can be much more efficient in semiconductor nanocrystals because of the relaxation in momentum conservation and a "forced" overlap of carrier wavefunctions induced by strong quantum confinement. For example, in 2-nm radius CdSe nanocrystals, the Auger decay time of a two e-h pair state (biexciton) is only 50 ps (ref. 7). The energy released in this process is approximately equal to the energy gap ($E_g \approx 2$ eV), and it is transferred to the remaining exciton, which corresponds to a heating rate of ca. 0.04 eV/ps per e-h pair. The latter value is close to the energy-loss rate typically observed in bulk CdSe and CdS (~0.1 eV/ps; ref. 12), indicating that in nanosized particles Auger heating can significantly alter carrier cooling dynamics at high pump intensities.

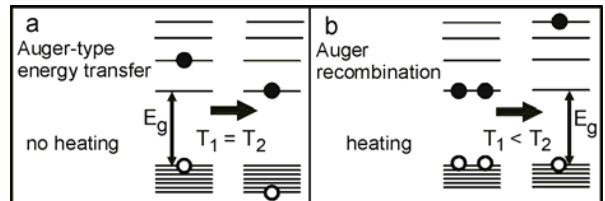

**Figure 1** Schematic representation of two different types of Auger effects in semiconductor nanocrystals. **a**, Auger-type e-h energy transfer does not change the average energy per e-h pair (i.e., carrier temperature). **b**, Auger recombination increases the total energy of carriers in a nanocrystal by approximately $E_g$ and, therefore, heats the electronic system.

In this article, we analyze the effect of Auger heating on energy relaxation dynamics in elongated CdSe nanocrystals [quantum rods (QRs)]. At high pump-intensities (more than 2-3 e-h pairs per QR), we detect a dramatic, order-of-magnitude reduction in the energy relaxation rate resulting from efficient Auger heating. The peculiarity of this regime is that the energy relaxation directly correlates with recombination dynamics, which has never been previously observed either in bulk or low-dimensional materials. Furthermore, we find that Auger heating differs in short and long QRs that can be explained by the difference in the scaling of Auger rates with respect to carrier density in zero-dimensional (0D) and 1D semiconductors[8].

In this work, we study highly monodisperse, colloidal CdSe QRs (Fig. 2a and b) prepared as hexane solutions[13-15]. We use a series of samples with the same QR diameter of 4.6 nm and various lengths from 22 to 44 nm. The samples are excited at 400 nm using frequency-doubled, 100-fs pulses from an amplified Ti:sapphire laser (100-kHz repetition rate). Time-resolved photoluminescence (PL) measurements are performed using a femtosecond PL up-conversion (uPL) technique[16] with a time resolution of approximately 300 fs. All measurements are conducted at room temperature. More details on experimental procedures can be found in Methods.

Figure 2c displays uPL spectra of 29 nm long QRs taken at different time delays ($\Delta t$) after excitation for the initial e-h density, $n_{eh}$, of ~$5.5\times10^{18}$ cm$^{-3}$ that corresponds to 2-3 e-h pairs per QR on average. In contrast to hot PL spectra of spherical nanocrystals[17,18] (quantum dots) that show well separated emission peaks arising from distinct quantized states, the time-resolved spectra of QRs consist of a single peak with an extended, high-energy tail, which reflects the distribution of charge carriers, $n(\hbar\omega)$, over the dense manifold of high-energy QR states. In addition to $n(\hbar\omega)$, the shape of the PL spectrum is determined by spectral distributions of the interband transition oscillator strength, $f(\hbar\omega)$, and the e-h joint density of states, $\rho(\hbar\omega)$: $I_{PL}(\hbar\omega) \propto f(\hbar\omega)\rho(\hbar\omega)n(\hbar\omega)$. Since the absorption coefficient, $\alpha_0(\hbar\omega)$, is proportional to $f(\hbar\omega)\rho(\hbar\omega)$, we can extract the population-related term by dividing the PL spectra by $\alpha_0(\hbar\omega)$. Applying this procedure to measured uPL spectra, we find that the high-energy tail of $n(\hbar\omega)$ can be well described by the exponential dependence $n(\hbar\omega) \propto \exp(-\hbar\omega/kT_e)$ (inset in Fig. 2c; $k$ is the Bolzmann's constant), implying that carrier distributions produced in our experiments are thermal. These observations further suggest that carrier thermalisation occurs on time scales that are shorter than our time resolution (~ 300 fs) and the carrier temperature, $T_e$, can be directly derived from high-energy tails of the uPL spectra.

A progressive increase of the high-energy slope of uPL spectra with time in Fig. 2c reflects carrier-cooling dynamics, which is displayed in Fig. 2d by solid squares. At pump levels below ~$5.5\times10^{18}$ cm$^{-3}$, the temperature relaxation (time constant is 0.5 ps) does not show a significant dependence on either rod length or pump level, indicating that high-carrier-density effects play a minor role in this intensity range.

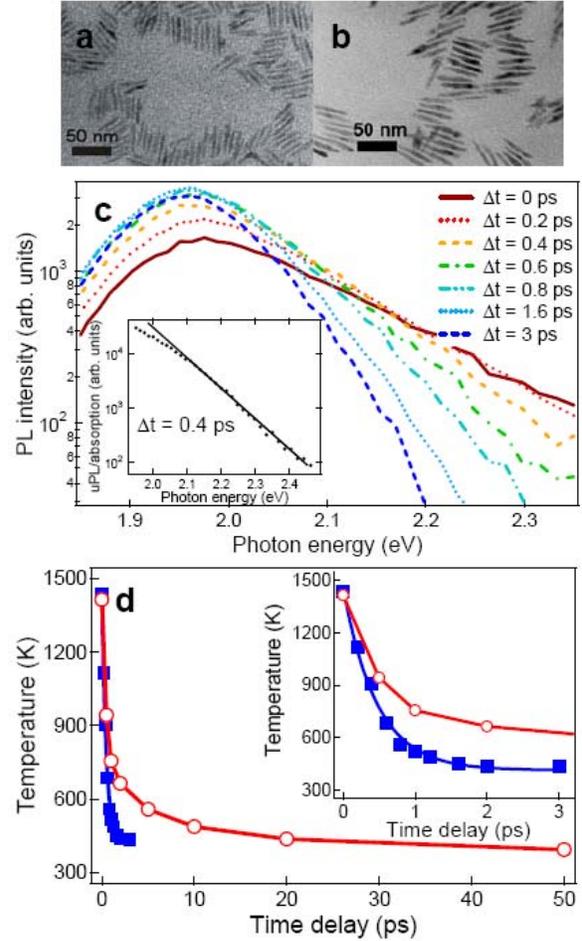

**Figure 2** Time-resolved uPL spectra of QR samples and carrier temperature dynamics derived from them. **a, b,** Examples of TEM images of two CdSe QR samples. The average rod length is 29 nm in **a** and 40 nm in **b**. The average QR diameter for both samples is 4.6 nm. **c,** Time-resolved uPL spectra of 29-nm long CdSe QRs measured at different time delays after excitation ($n_{eh}$ = $5.5\times10^{18}$ cm$^{-3}$). The inset shows an example of an exponential fit (line) to the high-energy tail of the population-distribution spectrum (solid circles) calculated as the ratio of the uPL spectrum measured at $\Delta t$ = 0.4 ps to the $\alpha_0(\hbar\omega)$ spectrum. **d,** Carrier temperature dynamics extracted from the uPL spectra measured at $n_{eh}$ = $5.5\times10^{18}$ cm$^{-3}$ (blue solid squares) and $n_{eh}$ = $2.2\times10^{19}$ cm$^{-3}$ (red open circles) for the same sample as in **c**. The inset is a blowup of the carrier temperature dynamics during the first 3 ps. The blue line is an exponential fit with a time constant of 0.5 ps.

In Fig. 3a, we plot the energy-loss rate [$\varepsilon_r = d(1.5kT_e)/dt$] as a function of $T_e$ (solid squares). The initial rate is approximately 0.2 eV/ps. It stays nearly constant until $T_e$ reaches ~700 K and then steeply drops by orders of magnitude as $T_e$ approaches the sample temperature. We find that the final electron temperature derived from the uPL data is always higher than 300 K (the nominal sample temperature) even at low excitation densities. This divergence is primarily due to the size dispersion of the QRs leading to $T_e$-independent broadening of the emission spectra. For example, for the QR sample in Fig. 2c, the slope of the high-energy tail of the PL spectrum measured using low-intensity, continuous-wave excitation (spectroscopic lamp) formally corresponds to the temperature of 370 K, which is close to the final carrier temperature (~400 K) measured for the same sample in the PL up-conversion studies.

Both the initial value of the energy relaxation rate measured at low pump intensities and its qualitative behaviour with changing carrier temperature are similar to those observed in bulk II-VI semiconductors at comparable excitation densities. In bulk semiconductors this behaviour has been explained in terms of strong coupling between the e-h and the LO-phonon subsystems that are in equilibrium with each other and cool together via interactions with acoustic phonons (see, e.g., refs. 19, 20). The temperature dependence of the relaxation rate in this regime can be described by the following expression[12]:

$$\varepsilon_r = \varepsilon_{ph} = -\frac{3}{2}\frac{\hbar\omega_0}{\tau_{LO}}\left(e^{\frac{\hbar\omega_0}{kT_a}} - e^{\frac{\hbar\omega_0}{kT_e}}\right)\frac{N_{LO}(T_a)}{N_{LO}(T_e)}\left(\frac{kT_e}{\hbar\omega_0}\right)^2 e^{\frac{\hbar\omega_0}{kT_e}}, \quad (1)$$

in which $\tau_{LO}$ is the characteristic LO-phonon decay time, $T_a$ is the acoustic phonon temperature and $N_{LO}(T)$ is the LO-phonon occupation number for $T$.

We can use equation (1) to accurately explain our results for energy relaxation rates measured at low pump intensity assuming that $T_a = 410$ K and $\tau_{LO} = 0.5$ ps (Fig. 3a, solid squares). This agreement provides strong evidence that at low pump levels, carrier intraband relaxation in QRs is bulk-like. This behaviour results from the high density of electronic states in QRs and is different from what is observed in quantum dots with sparse energy spectra [3,4,18].

Despite this similarity with bulk semiconductors, carrier relaxation in QRs shows several distinct features arising from the nanoscale size regime. One interesting observation is that the final carrier temperature in QRs (corrected for spectral broadening associated with sample polydispersity) is close to room temperature, while in bulk semiconductors, significant overheating is already observed at $n_{eh} = 10^{18}$ cm$^{-3}$ (ref. 12). In the bulk case, overheating occurs due to nonequilibrium filling of acoustic modes, which can eventually result in the situation for which $T_e = T_a$ and, hence, the cooling of the e-h system is controlled by the decay of the acoustic phonons (acoustic-phonon bottleneck[12]). The cooling of acoustic modes is controlled by relatively slow, heat-diffusion-limited interactions with the environment. The corresponding time constant ($\tau_a$) is determined by the diameter of the excitation spot and is typically in the µs to ms time range.

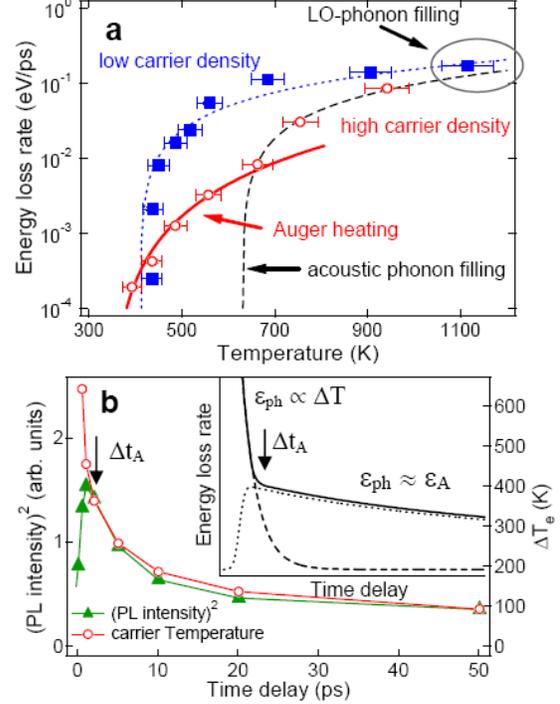

**Figure 3** Carrier cooling in the regime controlled by the competition between Auger heating and phonon emission. **a**, Energy loss rate $\varepsilon_r$ per carrier extracted from the cooling curves in Fig. 2 at low (5.5×10$^{18}$ cm$^{-3}$, blue solid squares) and high (2.2×10$^{19}$ cm$^{-3}$, red open circles) carrier densities. The blue dotted and the dashed black lines are calculated using the acoustic-phonon bottleneck model [Eq. (1)] for lattice temperatures of 410 K and 630 K, respectively. The red solid line is a fit to the quadratic dependence [$\varepsilon_r \propto (\Delta T_e)^2$] expected for cooling in the Auger regime. The error in $T_e$ is 5% and it is primarily determined by the accuracy of the exponential fit to the uPL spectra. The corresponding error in the calculated energy-loss rates is smaller than the symbol size. **b**, Time-dependent $\Delta T_e$ ($n_{eh} = 2.2\times10^{19}$ cm$^{-3}$) in 29-nm long QRs (red open circles) compared with the square of the spectrally integrated PL intensity (green solid triangles); the latter quantity is proportional to the square of the carrier density. Inset: Schematic illustration of time-dependent changes in the phonon energy loss rate $\varepsilon_{ph}$ at low (dashed line) and high (solid line) $n_{eh}$. In the latter case, $\varepsilon_{ph}$ decreases rapidly at short time delays, until at $\Delta t = \Delta t_A$ it reaches the Auger heating rate $\varepsilon_A$ (dotted line); starting from this point, $\varepsilon_{ph}$ closely follows $\varepsilon_A$.

In the case of QR samples, the efficiency of cooling of acoustic modes is increased by orders of magnitude because heat exchange is controlled by the dimensions of the individual nanoscale particle. We estimate that $\tau_a$ is on the sub-ps to ps time scale for QR dimensions studied in this paper. This time scale is comparable to that characteristic of the LO-phonon decay, which slows down the buildup of acoustic modes and hence significantly reduces the role of the acoustic-phonon bottleneck in nanosized rods. At high excitation densities energy relaxation changes significantly compared to that at low pump powers. Specifically, we find that in this case the cooling process is not terminated after a few ps (as it does at low pump intensities) but persists up to tens of ps [compare $T_e$ dynamics in Fig. 2d measured for $n_{eh} = 2.2\times10^{19}$ cm$^{-2}$ (open circles) and $5.5\times10^{18}$ cm$^{-2}$ (solid squares)]. We also observe a dramatic change in the $\varepsilon_r$ vs. $T_e$ dependence (Fig. 3a, open circles). Instead of a sharp drop in the range of intermediate temperatures observed at low pump intensities, $\varepsilon_r$ starts to decrease immediately with reducing $T_e$ and this reduction continues steadily until the final temperature is reached. Furthermore, we observe that for a given $T_e$ the magnitude of $\varepsilon_r$ is greatly reduced compared to the low-intensity situation. For example, at high excitation intensity, $\varepsilon_r$ at 700 K is 0.008 eV/ps while at low pump levels it is ca. 0.1 eV/ps. These observations cannot be explained by the acoustic-phonon-bottleneck model alone (dashed line in Fig. 3a) and indicate the existence of other mechanisms that become active at high excitation densities. Below we show that the observed behaviour can be accurately described if we account for Auger heating.

At pump levels used in these measurements carrier decay is dominated by non-radiative Auger recombination. The Auger recombination rate ($R_A$) is a function of the carrier density and hence of the number of e-h pairs per nanocrystal, $N$: $\left.\frac{dN}{dt}\right|_{Auger} = -R_A(N)$ [7,8].

The $N$-e-h-pair decay time constant is $\tau_N = N/R(N)$. Auger recombination leads to the heating of the electronic system by releasing energy that is approximately equal to the energy gap ($E_g$) in each recombination event. The Auger decay of the $N$ e-h pair state produces the $(N-1)$ state and, therefore, the corresponding heating rate ($\varepsilon_A$) per e-h pair is given by:

$$\varepsilon_A = \varepsilon_A(N) = E_g[\tau_N(N-1)]^{-1} = E_g R_A[N(N-1)]^{-1}. \quad (2)$$

During the initial fast cooling ($\Delta t < 1$ ps), energy relaxation is dominated by interactions with phonons with a rate that depends primarily on carrier temperature[12] and is only weakly dependent on e-h pair density through the weak effect of the acoustic-phonon bottleneck: $\varepsilon_{ph} = \varepsilon_{ph}(T_e,N)$. The phonon-related relaxation rate decreases with decreasing $T_e$ and eventually becomes equal to the Auger heating rate, which marks the onset of the relaxation stage in which carrier intraband dynamics are controlled by the Auger process (inset of Fig. 3b). Starting from this point ($t = \Delta t_A$) the e-h temperature is determined by the equilibrium between the energy outflow through interactions with phonons and the energy inflow because of Auger heating:

$$\varepsilon_{ph}(T_e, N) = \varepsilon_A(N). \quad (3)$$

This equilibrium is maintained by the *negative-feedback* mechanism, which operates in the following way. If $\varepsilon_{ph}$ drops below $\varepsilon_A$, the Auger heating takes over, which increases $T_e$ and hence $\varepsilon_{ph}$ [see equation (1)]. In the opposite case of $\varepsilon_{ph} > \varepsilon_A$, carrier cooling due to electron-phonon interactions dominates over Auger heating, which drives $T_e$ down until $\varepsilon_{ph}$ becomes equal to $\varepsilon_A$. The above considerations imply that during the Auger-process-controlled relaxation stage, $T_e$ is a function of carrier density and it can be determined from equation (3). The direct correspondence between carrier density and $T_e$ is clearly manifested in both measured cooling dynamics and pump-intensity dependent data as discussed below.

In our pump-intensity-dependence studies, we concentrate on two QR samples with two different QR lengths of 22 nm (short rods) and 44 nm (long rods). According to our previous studies, these two samples show distinctly different carrier-density dependences of Auger recombination rates[8]. In short rods that can be considered as 0D objects carriers are present in the form of unbound e-h pairs and, therefore, Auger rates are cubic with respect to carrier density: $R_A(N) \propto N^3$. In the case of long, 1D rods, electrons and holes are bound into 1D excitons that recombine in the bimolecular fashion with $R_A(N) \propto N^2$. This difference in scaling of $R_A$ should have a pronounced effect on the pump-dependence of carrier overheating $\Delta T_e = T_e - T_L$ (where $T_L$ is the nominal lattice temperature), if it is indeed due to the Auger process.

If we neglect a weak dependence of $\varepsilon_{ph}$ on carrier density, we obtain that at moderate overheating [$\Delta T_e \hbar \omega_{LO}/(kT_eT_a) < 1$, $T_a \approx T_L$], the phonon-related relaxation rate is approximately proportional to $\Delta T_e$. From the condition of the equilibrium between $\varepsilon_{ph}$ and $\varepsilon_A$ [equation (3)], we obtain $\Delta T_e(N) \propto \varepsilon_A(N)$. Finally using equation (2), we find that $\Delta T_e(N)$ is proportional to $R_A(N)$ and, hence, the pump-power dependence of the ratio of the overheating measured in 0D and 1D rods scales as $N$: $\Delta T_e(N)|_{0D}/\Delta T_e(N)|_{1D} \propto N$. To account for the dependence of $\varepsilon_{ph}$ on the carrier density, we need to take into consideration the acoustic-phonon bottleneck. Because of extremely efficient heat exchange between nanoscale QRs and the outside medium (solvent in this case), the non-equilibrium heating of acoustic modes in QRs is not significant and, therefore, it can be accounted for in the linear approximation: $T_a - T_L \propto N$ (ref. 12). The use of a corrected expression for $\varepsilon_{ph}$ in equation (3) modifies the dependence of $\Delta T_e$ on $N$; however, it still preserves an approximate linearity of the ratio $\Delta T_e(N)|_{0D}/\Delta T_e(N)|_{1D}$ with respect to $N$.

To experimentally compare carrier heating behaviour in 0D and 1D cases, we analyze the carrier density dependence of $\Delta T_e$ measured at $\Delta t = 2$ ps (i.e., after the equilibrium between $\varepsilon_{ph}$ and $\varepsilon_A$ is established; see below) for 22 nm and 44 nm long QRs that are characterized by cubic and quadratic Auger recombination rates, respectively[8]. As a measure of carrier density we use an average number of e-h pairs per QR, $\langle N \rangle$, which is calculated as $\langle N \rangle = j_p \sigma(\hbar\omega_p)$, where $j_p$ is the pump per-pulse fluence (in photons per cm$^2$) and $\sigma(\hbar\omega_p)$ is the QR absorption cross section at the pump photon energy, $\hbar\omega_p$. We clearly observe that shorter rods show a faster increase of the temperature with increasing pump level ($\Delta T_e \propto \langle N \rangle^{1.8}$) than longer rods ($\Delta T_e \propto \langle N \rangle^{0.9}$) (Fig. 4). Furthermore, the ratio of $\Delta T_e$ measured for these two samples scales as $\langle N \rangle^{0.9}$, which is in a good agreement with the prediction of the Auger-heating model.

These results are also consistent with observed cooling dynamics. In Fig. 3b, we compare relaxation of $\Delta T_e$ with dynamics of PL intensity squared, $I_{PL}^2$, and observe that both transients show nearly identical behaviours at delay times $\Delta t > 2$ ps. Since $I_{PL}$ is proportional to the carrier density ($I_{PL} \propto n_{eh}\tau_r^{-1}$; $\tau_r$ is the radiative decay time constant), the above observation indicates that $\Delta T_e$ is proportional to $\langle N \rangle^2$, which agrees with results of the pump-intensity-dependence studies (short rod case in Fig. 4a). The dynamical data also provide information on the onset ($\Delta t_A$) of the Auger-controlled cooling stage, which corresponds to the time delay, after which $\Delta T_e$ and $I_{PL}$ show "correlated" relaxation (~2 ps in Fig. 3b).

Finally, we demonstrate that the Auger-heating model also allows us to accurately describe the temperature dependence of the energy-loss rates measured at high pump fluences. In the Auger heating regime, $\varepsilon_r \propto [d(\Delta T_e)/dN](dN/dt) = [d(\Delta T_e)/dN]R_A$. Our experimental data in Fig. 4 indicate that $\Delta T_e$ is approximately proportional to $\langle N \rangle^2$ and $\langle N \rangle$ in short and long rods, respectively, which yields $\varepsilon_r \propto (\Delta T_e)^2$ for both cases. The measured temperature dynamics indicate that the Auger-heating regime establishes at temperatures below ca. 700 K. In this range the quadratic dependence on $\Delta T_e$ describes the behaviour of the energy loss rate (solid red line in Fig. 3b) remarkably well.

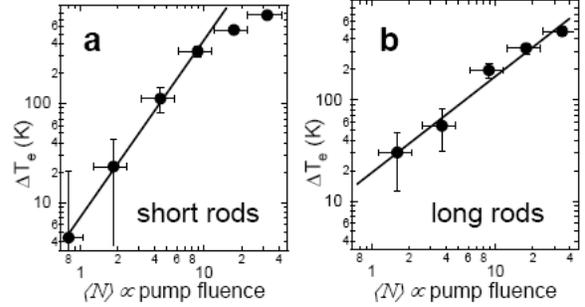

**Figure 4** Carrier overheating, $\Delta T_e$, at $\Delta t = 2$ ps as a function of the average number of e-h pairs per QR. **a** and **b**, 22-nm and 44-nm long rods that can be treated as 0D and 1D objects, respectively. The error in $\Delta T_e$ results from the 5% error in $T_e$ as explained in Fig. 3a. The error in $\langle N \rangle$ is primarily due to an estimated 30% uncertainty in QR absorption cross sections. These data are fit to the dependence $\Delta T_e \propto \langle N \rangle^\beta$, where $\beta = 1.8$ and 0.9 in **a** and **b**, respectively. The deviation from this power dependence for shorter rods observed at high pump intensities is likely a result of a very fast shortening of Auger recombination times, $\tau_A$, with increasing number of e-h pairs per QR. The fast Auger decay leads to an appreciable reduction in $\langle N \rangle$ during the first 2 ps compared to its nominal value derived from the pump fluence (plotted along the horizontal axis). This effect is more pronounced for short, 0D QRs, for which $\tau_A$ scales faster with $N$ than for long, 1D rods ($N^{-2}$ vs $N^{-1}$, respectively).

In conclusion, we study carrier cooling dynamics in semiconductor QRs as a function of rod length and pump intensity. We observe a significant reduction of the energy-loss rate (by more than one order of magnitude) at high pump levels resulting from highly efficient Auger heating. In this regime, the carrier temperature is determined by the balance between energy outflow through interactions with phonons and energy inflow produced by Auger recombination. This approximate equilibrium results in a cooling dynamics that is controlled by the carrier recombination process. Because of the direct correlation between energy-relaxation and recombination dynamics, the carrier cooling behaviour is significantly different in short and long QRs due to the difference in scaling of Auger recombination rates with respect to carrier density in 0D and 1D semiconductors.

## METHODS

We fabricated QR samples using colloidal synthetic procedures adapted from refs. 13 – 15. In addition to trioctylphosphine (TOP) and trioctylphosphine oxide (TOPO), standard coordinating ligands used in the preparation of spherical CdSe nanocrystals[13], we used a combination of hexylphosphonic acid (HPA) and tetradecylphosphonic acid (TDPA) to promote formation of elongated, rod-shaped particles. Typically, a particular HPA:TDPA ratio (e.g., 1:1 to 1:3) was chosen to achieve a desired rod diameter, while different rod lengths were obtained by selecting aliquots from the reaction mixture over time. Longer reaction times yielded longer rods, with little impact on rod diameter. This procedure produced highly monodisperse QR samples that had 5%–7% and 10% standard deviations for QR radii and lengths, respectively.

Steady-state PL quantum yields of QRs are a few percent, which is typical for samples fabricated without subsequent overcoating with an inorganic layer of a wide-gap semiconductor. These relatively low quantum yields imply a significant role of nonradiative recombination processes in our samples. Based on typical room-temperature radiative lifetimes for QRs (several nanoseconds to tens of nanoseconds), we can estimate that the average nonradiative time constants in QR samples are on the hundreds of picosecond timescales. Since these timescales are significantly longer than those of cooling processes studied in this work (sub-picosecond to tens of picosecond time constants), the results of our relaxation studies are not significantly affected by nonradiative decay channels and, specifically, by exact values of steady-state emission quantum yields.

In the PL up-conversion experiment[16], the QR samples were excited at 400 nm with 100-fs pulses from a frequency-doubled, amplified Ti:sapphire laser (100-kHz repetition rate). An elliptical mirror was used to collect and to re-focus the QR PL onto a nonlinear-optical $\beta$-barium borate (BBO) crystal. The sample emission was frequency-mixed (gated) in the BBO crystal with variably delayed 100-fs pulses at a fundamental frequency of the Ti-sapphire laser. The up-converted, sum-frequency signal was spectrally filtered by a monochromator and detected with a cooled photomultiplier tube coupled to a photon counting system. By scanning the monochromator and the time delay between excitation and gating pulses we were able to obtain spectrally and time-resolved PL data. The spectral resolution in these measurements was ~5 nm and the temporal resolution was ~300 fs.

## Acknowledgements

The authors thank Dr. H. Htoon for useful discussions. This work was supported by Los Alamos LDRD Funds and the Chemical Sciences, Biosciences, and Geosciences Division of the Office of Basic Energy Sciences, Office of Science, U.S. Department of Energy. Correspondence and requests for materials should be addressed to V.I.K.

## Competing financial interests

The authors declare that they have no competing financial interests.